# Pyroelectric doping reversal of MoS$_2$ p-n junctions on ferroelectric domain walls probed by photoluminescence

Javier Fernández-Martínez[1,2], Joan J. Ronquillo[1], Guillermo López-Polín[1,3], Herko P. van der Meulen[1,2,3], Mariola O Ramírez[1,2,3] and Luisa E. Bausá[1,2,3]

[1]Dept. Física de Materiales, Universidad Autónoma de Madrid, 28049-Madrid, Spain

[2]Instituto de Materiales Nicolás Cabrera (INC), Universidad Autónoma de Madrid, 28049-Madrid, Spain

[3]Condensed Matter Physics Center (IFIMAC), Universidad Autónoma de Madrid, 28049-Madrid, Spain

**Abstract**

Tailoring the optical properties and electronic doping in transition metal dichalcogenides (TMDs) is a central strategy for developing innovative systems with tunable characteristics. In this context, pyroelectric materials, which hold the capacity for charge generation when subjected to temperature changes, offer a promising route for this modulation.

This work employs spatially resolved photoluminescence (PL) to explore the impact of pyroelectricity on the electronic doping of monolayer MoS$_2$ deposited on periodically poled LiNbO$_3$ (LN) substrates. The results demonstrate that pyroelectricity in LN modulates the charge carrier density in MoS$_2$ on ferroelectric surfaces acting as doping mechanism without the need for gating electrodes.

Furthermore, upon cooling, pyroelectric charges effectively reverse the doping of p-n junctions on DWs, converting them into n-p junctions. These findings highlight the potential of pyroelectric substrates for tunable and configurable charge engineering in transition metal dichalcogenides and suggest their applicability to other combinations of 2D materials and ferroelectric substrates. They also open avenues for alternative device architectures in nanoelectronic or nanophotonic devices including switches, memories or sensors.

## 1. Introduction

Among the family of two-dimensional (2D) materials, monolayer (1L) transition metal dichalcogenides (TMDs) are nowadays in the spotlight due to their unique electronic, optical and mechanical properties.[1] 1L-TMDs combine the potential for novel optoelectronic devices, such as nanometrically thin light sources[2-4] or flexible electronic components,[5,6] with the possibility of exploring quantum-derived new properties and effects. Namely, their direct band gap in the visible range of the spectrum enables them to strongly interact with light, while their large exciton binding energy, of the order of several tens of meV, permits the existence of quasiparticles such as excitons and trions at RT.[7] In addition, their atomical thickness, smaller than 1 nm, offers a unique opportunity to tune their characteristics by means of their surrounding environment through mechanical strain[8] or chemical doping.[9] In this context, the association of 1L-TMDs with ferroelectric substrates has emerged as a promising approach to provide 1L-TMDs with novel properties without the need for complex fabrication processes, which could hinder their implementation in optoelectronic devices. A relevant example of such a type of substrates is lithium niobate $LiNbO_3$ (hereafter LN), one of the most extensively used dielectric materials for optoelectronics. Research in photonics and material science has benefited from its high nonlinear coefficients for second and third order processes, its strong intrinsic electric polarization, and elevated electro-optical, piezoelectric and pyroelectric coefficients.[10] In fact, LN stands out as a highly versatile multifunctional platform, offering significant potential for advanced nanophotonic applications.[11] A recent advance involves the integration of LN with 1L-$MoS_2$ to achieve pulsed laser operation at the nanoscale.[12]

The synergy between 2D materials and LN has been previously addressed, revealing their potential for next-generation photonic devices and cutting-edge technologies. Among the earliest investigations into 2D/LN heterostructures, the integration of graphene with LN unveiled phenomena such as domain-dependent electrostatic doping[13] and persistent photogating effects.[14] Subsequent research on TMD/LN heterostructures has taken advantage of the direct bandgap and luminescence properties of monolayer 1L-TMDs. Notably, optical control of excitonic quasiparticles, including excitons and trions, has been demonstrated in $MoSe_2$ and $WSe_2$ deposited on LN.[15,16]

In a recent work, the combined effect of ferroelectricity and light on the electron doping and optoelectronic properties of 1L-$MoS_2$, deposited on periodically poled lithium niobate (PPLN) was studied.[17] The results highlighted the presence of a ferroelectrically-induced electron photodoping process which depends on the direction of the spontaneous polarization. Further, p-n homojunctions in the vicinities of the ferroelectric domain walls (DWs) were formed. Their optical characterization unveiled the presence of an intense out-of-plane electric field in the DW surface, which strongly modulate the photoluminescence (PL) of 1L-$MoS_2$.

Here, we use PL spectroscopy to demonstrate the ability of pyroelectricity to control the electronic doping of 1L-$MoS_2$, providing a simple, electrode-free approach that poses a challenge in these systems. Pyroelectricity of LN has been widely used in relevant applications, such as particle trapping,[18,19] surface charge-assisted lithography,[20] and enhancing the sensitivity of 2D material-based photodetectors.[21,22] However, its role in modulating the electronic doping of 2D materials across a broad temperature range remains little studied.[15, 23]

In this work, we exploit the strong sensitivity of monolayer TMDs to their surrounding environment to investigate the modulation of the electron doping of 1L- $MoS_2$, deposited on the polar surface of a LN crystal as the temperature is directly reduced from 300 K to 10 K. Specifically, a LN with a hexagonal alternating ferroelectric domain structure was used as a substrate. This enables the study of the electron density modulation of 1L-$MoS_2$ on each type of ferroelectric 180º domain, by monitoring the trion-to exciton emission ratio in 1L-MoS2, which correlates with the free carrier density in the layer. The effect of pyroelectricity on electron doping was confirmed using $MoS_2$ flakes derived from two distinct bulk $MoS_2$ crystals, each with a different pristine background doping level.

By comparing the PL spectra at 300 and 10 K we unveil a pronounced change in the electronic doping of 1L-MoS$_2$. This change arises from the variation in the substrate's spontaneous polarization as the temperature decreases from 300 K to 10 K and reflects the modification in the balance between polarization and screening charges at the MoS$_2$/LN interface. Moreover, the resulting change in the charge density, attributable to the pyroelectric effect of LN, is similar for both types of MoS$_2$ monolayers and agrees with the reported values for the variation in the spontaneous polarization in LN.

Finally, the impact of the domain wall on the spectral features of the 1L-MoS$_2$ emission has been analyzed at 10 K. Cryogenic temperatures drastically reduce electron-phonon interactions, leading to narrower linewidths compared to room temperature. This enables a more precise analysis of the evolution of exciton and trion on the vicinity of the domain walls. Notably, scanning micro-PL across the domain walls reveals the role of pyroelectricity in the switching of p-n to n-p junctions at the nanometric scale.

Our findings underscore the role of pyroelectricity in modulating the PL of quasiparticles in MoS$_2$ and demonstrate that ferroelectric substrates offer an effective strategy for precise control of electronic doping in TMDs. Our results open new avenues for engineering doping profiles in 2D materials, enhancing their potential in optoelectronic applications.

## 2. Results and discussion

**Figure 1**a presents a schematic of the 1L-MoS$_2$/LN sample, illustrating the MoS$_2$ monolayer and the ferroelectric domain-engineered LN substrate. Figure 1b shows an optical micrograph of a 1L-MoS$_2$ flake (S1), mechanically exfoliated and transferred onto the polar surface of a micropatterned LN crystal with alternating domain regions fabricated by DEBW. For clarity, the monolayer contour is outlined in the micrograph. The inverted hexagonal P$_{up}$ domains are highlighted against the P$_{down}$ background. They have a width of around 5 μm and are separated by distances of approximately 10 μm. Accordingly, monolayer flakes larger than 15-20 μm were selected to ensure overlap with both P$_{up}$ and P$_{down}$ regions.

Figure 1c shows representative room temperature emission spectra of 1L-MoS$_2$ on domains with opposite polarity under intense light illumination (6.6×10$^5$ W/cm$^2$). The emission spectra were obtained in the spectral region of the A exciton (630 - 710 nm emission range). Despite the intensity difference, both spectra exhibit an asymmetric shape due to the contribution of the emission from neutral excitons and negatively charged excitons (trions). Henceforth, we will refer to A excitons as X (peak position at 655 nm, 1.89 eV) and trions as X$^-$ (peak position at 677 nm, 1.83 eV). As seen, the relative contribution of these quasiparticles to the emission spectra strongly depends on the orientation of the underneath spontaneous polarization. Specifically, the emission from 1L-MoS$_2$ on the P$_{down}$ domain is primarily dominated by excitons, while the trion emission governs the PL properties of 1L-MoS$_2$ on P$_{up}$ domains. Additionally, since at room temperature trions exhibit a lower quantum yield than excitons,[24] the overall 1L-MoS$_2$ PL intensity on P$_{up}$ domains, where trions dominate, is reduced. This is further illustrated in Figure 1d, which shows the spatial distribution of the integrated PL intensity in the 640-710 nm range. The lower-intensity regions associated with P$_{up}$ polarization, and the higher-intensity regions associated with P$_{down}$ domains -where exciton emission predominates- reveal a very good correlation between the ferroelectric patterning of the substrate (Fig. 1b) and the PL intensity of 1L-MoS$_2$. As explained below, the relative contribution of trions and excitons to the emission spectra will be used in our work to determine the electron doping in 1L-MoS$_2$ on each type of ferroelectric domain.

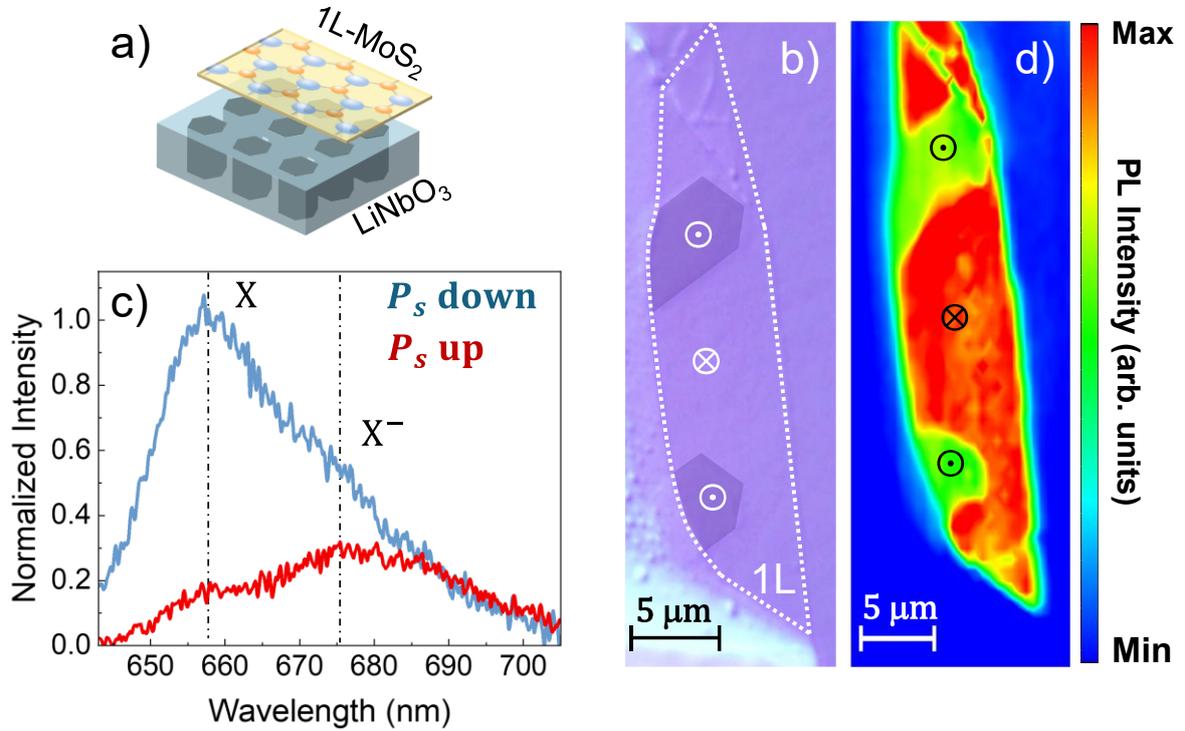

**Figure 1**. a) Schematics of the transferred MoS$_2$ monolayer on top of a domain-engineered LN crystal. The dark grey hexagonal prisms represent upward-polarized domains within a crystal background with downward polarization in light grey. b) Optical micrograph of a 1L-MoS$_2$ (sample S1) transferred onto the surface of the domain patterned LN substrate. The contrast between regions with opposite ferroelectric polarization is enhanced using false-colour imaging. The contour of the monolayer and the orientation of the spontaneous polarization of the substrate are indicated (⊙ for P$_{up}$, ⊗ for P$_{down}$). c) Room temperature PL spectra of 1L-MoS$_2$ obtained on the P$_{down}$ (red) and P$_{up}$ (blue) ferroelectric domains. The spectral positions of the A exciton (X) and trion (X$^-$) are marked with dashed lines. d) Spatial distribution of the integrated PL intensity. The orientation of the ferroelectric polarization is indicated.

An important point to address is the interplay between light and ferroelectricity in this system. As previously demonstrated, the charge carrier density in 1L-MoS$_2$ exhibits a dependence on the ferroelectric domain orientation only at high excitation densities. Specifically, as reported by the authors, the difference in the charge density is associated with a photoinduced charging process that requires intense light illumination.[17] To illustrate this domain-dependent photodoping process, **Figures 2**a and 2b compare the normalized PL spectra of 1L-MoS$_2$ (S1) obtained on domains with opposite polarization orientation for excitation intensities in the order of 10$^5$ W/cm$^2$ and 10$^3$ W/cm$^2$, respectively. The deconvoluted spectra for the exciton (blue) and trion (red) are presented for each domain orientation. As observed in Figure 2a, under high excitation intensities (6.6×10$^5$ W/cm$^2$), the charge density of the monolayer is modulated by the polarization orientation of the substrate, favoring the dominant presence of negatively charged excitons (trions) on the surface of P$_{up}$ domains due to an increase in electron density compared to the P$_{down}$ domains.

However, at lower excitation intensity (8×10$^3$ W/cm$^2$) the shape of the PL spectra and the trion-to-exciton intensity ratio are similar for both P$_{down}$ and P$_{up}$ domains (see Figure 2b), indicating a negligible difference in the charge carrier density of 1L-MoS$_2$ between both types of domains. This implies a minimal contribution of domain-dependent photodoping, in agreement with previous results for 1L-MoS$_2$ on LN at moderate-low excitation intensity.[17]

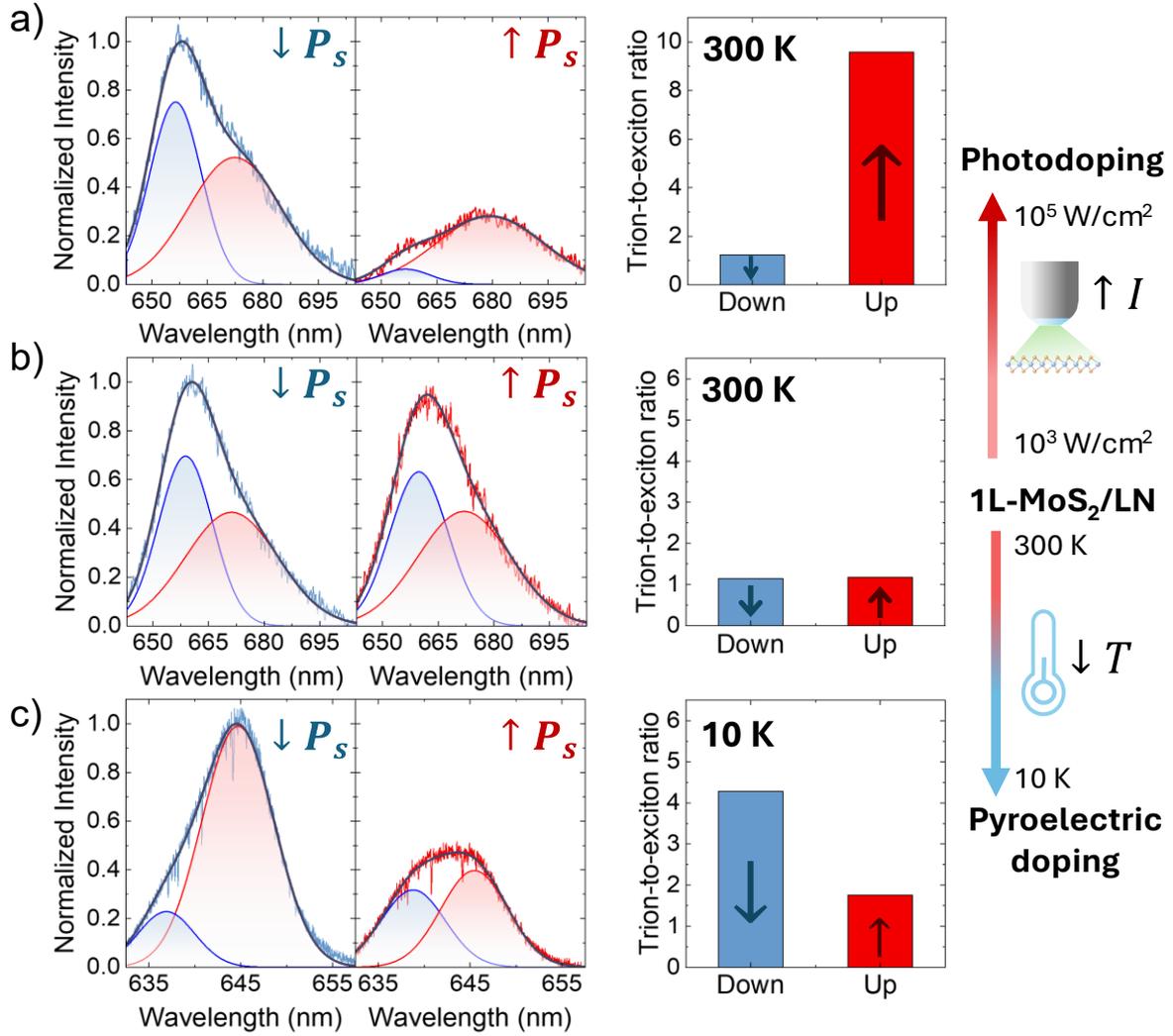

**Figure 2**. Deconvoluted PL spectra of the S1 1L-MoS$_2$ sample on top of a P$_{down}$ and P$_{up}$ domains at 300 K (a) under photodoping conditions and (b) in the absence of photodoping. c) Deconvoluted PL spectra obtained upon cooling to 10 K for the P$_{down}$ and P$_{up}$ domains in the absence of photodoping. For each case, the relative trion-to-exciton intensity ratio in the P$_{down}$ and P$_{up}$ domains are shown in blue and red, respectively, in the right panels.

The electron doping level of the MoS$_2$ monolayer in each domain can be estimated by analyzing the trion-to-exciton intensity ratio ($I_{X^-}/I_X$) in the emission spectra. Considering the mass action law, which governs the equilibrium between excitons, trions, and free electrons, the electron density ($n_e$) can be estimated from the PL spectra without the need for electrical measurements, as follows: [9]

$$\frac{I_{X^-}}{I_X} = \frac{n_e}{\eta_r C(T)}, \quad (1)$$

where $I_{X^-}$ and $I_X$ represent the intensity contribution of trions and excitons to the spectra, respectively. $\eta_r$ denotes the relative quantum yield of the exciton and the trion ($\eta_r = \eta_X/\eta_{X^-}$), while $C(T)$ is a function of temperature, which is given by

$$C(T) = \left(\frac{4 m_X m_e}{\pi \hbar^2 m_{X^-}}\right) \cdot k_B T \cdot e^{-E_b/k_B T}, \quad (2)$$

Here, $m_X$, $m_{X^-}$ and $m_e$ refer to the effective masses of excitons, trions and electrons, respectively, with values of $m_X = 0.8 m_0$, $m_{X^-} = 1.15 m_0$ and $m_e = 0.35 m_0$ for 1L-MoS$_2$,[25] where $m_0$ is the free electron mass. The trion binding energy is $E_b \simeq 20$ meV.[7] At 300 K, $C(T)$ takes a value of 4.86·10$^{12}$ cm$^{-2}$, while $\eta_r$ is approximately 20/3.[9]

On this basis, and considering the trion-to-exciton ratios shown in Figure 2, the electronic density at low excitation intensity is similar to that of pristine 1L-MoS$_2$ with an estimated value of $n_e^i$ = (3.8 ± 0.5)×10$^{13}$ cm$^{-2}$, where the superscript stands for "intrinsic". As the excitation intensity is increased, this value evolves to (3.2 ± 0.05)×10$^{14}$ cm$^{-2}$ and (4.0 ± 0.5) ×10$^{13}$ cm$^{-2}$ for the MoS$_2$ monolayer on the P$_{up}$ and P$_{down}$ domains, respectively, in excellent agreement with the previously reported polarization-mediated selective photodoping at high excitation intensities.[17]

The results presented so far confirm the capability of LN to modulate the electron density of 1L-MoS$_2$ by means of a ferroelectric-driven photoinduced charging process, in which light induces charge generation and transfer at the 1L-MoS$_2$/LN interface interface.

In this context, once the presence of photodoping has been established and analyzed, from here on, we will employ conditions to eliminate this effect and isolate the impact of pyroelectricity on the modulation of emission and doping. In fact, LN pyroelectricity emerges as an alternative tool to modulate the charge density at the 1L-MoS$_2$/LN interface without the need for intense light illumination. Note that the modification in the value of the spontaneous polarization induced by temperature change would generate a modification in the balance between the polarization and screening charges at the 1L-MoS$_2$/LN interface, thereby leading to the modulation of the electronic doping in the flake.

To assess the influence of the pyroelectric effect of the substrate on the modulation of electron doping in 1L-MoS$_2$, the micro-luminescence spectra were recorded upon cooling to 10 K for each type of domain and compared to those at room temperature. For these experiments, the absence of photodoping at the employed excitation intensity was confirmed. The results are displayed in Figure 2c. From the deconvolution, the spectral positions of excitons and trions were found to be located at around 638 nm (1.94 eV) and 649 nm (1.91 eV), respectively, in agreement with previous results.[26] As compared to RT, the 10 K spectra exhibit a clear blueshift consistent with the increase in the bandgap due to lattice contraction of the monolayer.[26] A clear narrowing of the emission bands is also observed due to the reduced phonon population at low temperature. In addition, an increased trionic contribution is observed in both spectra, in agreement with the higher quantum yield of trions at low temperature.[27] However, unlike the room-temperature emission shown in Fig. 2b, upon cooling to 10 K the trion-to-exciton ratio in the spectra is noticeably different on each domain. Specifically, the 10 K PL emission of MoS$_2$ reveals an intense trionic contribution on the P$_{down}$ ferroelectric domain, which is approximately twice as high as that observed on the P$_{up}$ domain. This behavior indicates the presence of a different electron doping level in 1L-MoS$_2$ on each domain in contrast to what is observed at room temperature. The result suggests that the temperature-dependent variation in the spontaneous polarization P$_s$, that is, the pyroelectric effect, is responsible for the observed spectral differences between the two domains. In fact, the results are consistent with a modification of different polarization/screening-charge balance at the 1L-MoS$_2$/LN interface driven by the increase in the P$_s$ of the substrate as the temperature decreases from 300 K to 10 K. As expected from the increase in P$_s$, the results reveal a higher electron density in MoS$_2$ on the P$_{down}$ domain (with a significantly higher trion-to-exciton ratio) compared to the P$_{up}$ domain (which exhibits a lower trion-to-exciton ratio). This is consistent with the expected change in the charge balance at the interface.

In fact, the domain selective spatial modulation of the charge density upon decreasing the temperature from 300 to 10 K can be directly correlated with the change in the spontaneous polarization value of the substrate. Since the variation of P$_s$ is identical in magnitude but opposite in sign for both domains, the pyroelectric effect is assumed to symmetrically enhance or reduce the electron doping of the MoS$_2$ monolayer on each domain by the same amount relative to its value measured at room temperature. The variation in the charge density of 1L-MoS$_2$ on each domain, Δn, can be estimated by considering again its proportionality to the trion-to-exciton contribution in the spectra. At cryogenic temperature, the relationship between the electron charge densities on each domain is given by

$$\frac{n_e^\downarrow(10\text{ K})}{n_e^\uparrow(10\text{ K})} = \frac{n_e^i(300\text{ K}) + \Delta n}{n_e^i(300\text{ K}) - \Delta n} = \frac{\left(\frac{I_{X-}}{I_X}\right)_\downarrow^{10\text{ K}}}{\left(\frac{I_{X-}}{I_X}\right)_\uparrow^{10\text{ K}}} \qquad (3)$$

where $n_e^\downarrow(10\text{ K})$ and $n_e^\uparrow(10\text{ K})$ stand for the electron density of 1L-MoS$_2$ at 10 K on the P$_{down}$ and P$_{up}$ domain, respectively. $n_e^i(300\text{ K})$ represents that value at room temperature, being $n_e^i(300\text{ K})$ = 3.8×10$^{13}$ cm$^{-2}$, according to the previously obtained intrinsic electron density in the absence of photodoping. The relationship used above allows cancelling the ratio η$_r$ (see eq.1), which is altered at low temperature due to the variations in the radiative rates of trions and excitons.

From the deconvoluted PL spectra in Figure 2c, the electron charge density variation with respect to room temperature is obtained as $\Delta n \simeq$ (1.6 ± 0.4)×10$^{13}$ cm$^{-2}$, i.e. 2.6 ± 0.7 μC/cm$^2$. This value closely matches the reported variation in P$_s$ of LN when decreasing the temperature in the range 300-10 K, around 3 μC/cm$^2$,[28] confirming the pyroelectric effect as the primary cause of the observed charge modulation. The results are consistent with doping changes of the same order of magnitude as those recently reported in hBN-encapsulated graphene on LN as the temperature decreases down to 10 K.[23] The results also highlight the sensitivity of the PL of 1L-MoS$_2$ as a temperature and charge sensor capable to detect changes in the surrounding environment.

To further stress this point, additional experiments were performed on a different 1L-MoS$_2$ flake (sample S2) with a much larger background electron density.

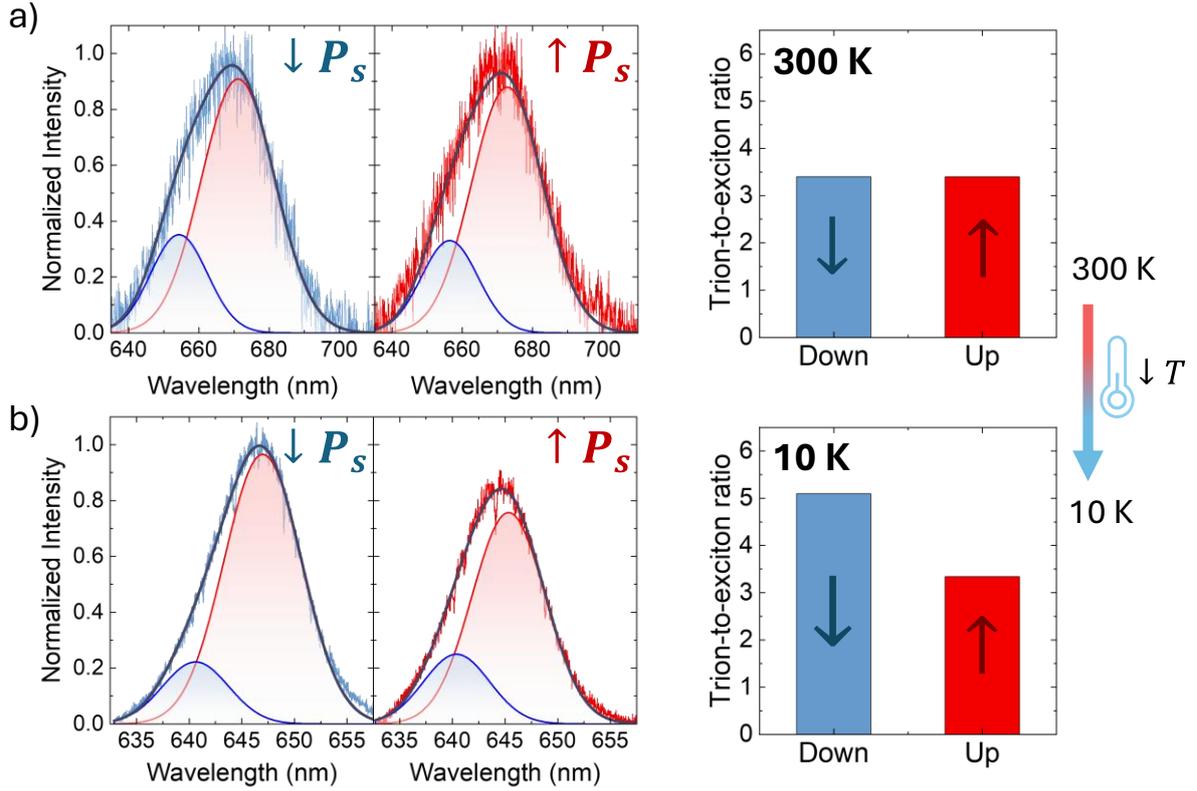

**Figure 3.** a) Deconvoluted PL spectra of the S2 1L-MoS$_2$ sample on the P$_{down}$ (left) and P$_{up}$ (right) ferroelectric domains at 300 K. The exciton and trion contributions are depicted in blue and red, respectively. b) Deconvoluted PL spectra of sample S2 measured upon cooling to 10 K. The relative trion-to-exciton contributions in the P$_{down}$ and P$_{up}$ domains are shown in blue and red, respectively, in the right panels.

**Figure 3**a shows representative normalized PL spectra measured on sample S2 at room temperature on the P$_{down}$ (left) and P$_{up}$ (right) domains. The spectra exhibit similar line shapes, characterized by a prominent trion peak and a less intense excitonic contribution. The trion-to-exciton ratios are comparable in both domains (see right panel), which according to equation 1, correspond to an electron density of approximately (1.10 ± 0.05)×10$^{14}$ cm$^{-2}$ on both domains, a doping level higher than that of sample S1. Figure 3b displays the 1L-MoS$_2$ normalized PL spectra at 10 K on each domain. At low temperature, the effect of the enhanced spontaneous polarization of the substrate is once again evident in the PL of the monolayer. At 10 K, in addition to the aforementioned changes in the position and width of the emission bands, the spectra obtained exhibit different relative contributions from excitons and trions on each domain, differing from the trend observed at room temperature. Namely, on the P$_{down}$ domain the increase in P$_s$ effectively increases the doping level of 1L-MoS$_2$. For sample S2, the variation in the charge density of the MoS$_2$ monolayer from 300 to 10 K has been determined from the spectra in Figure 3, taking into account Equation 3 and considering an intrinsic electron density of $n_e^i = 1.10 \times 10^{14}$ cm$^{-2}$, yielding Δn ≃ 3.5 ± 0.7 µC/cm$^2$. This value is similar to that obtained for sample S1, within the error margin, and corresponds to the change in the spontaneous polarization of the substrate from 300 to 10 K. Our findings underscore the ability of pyroelectricity to modulate the electronic doping of 1L-MoS$_2$. Moreover, the results indicate that the charge modulation induced by the pyroelectric effect is independent of the intrinsic doping within the analyzed doping range.

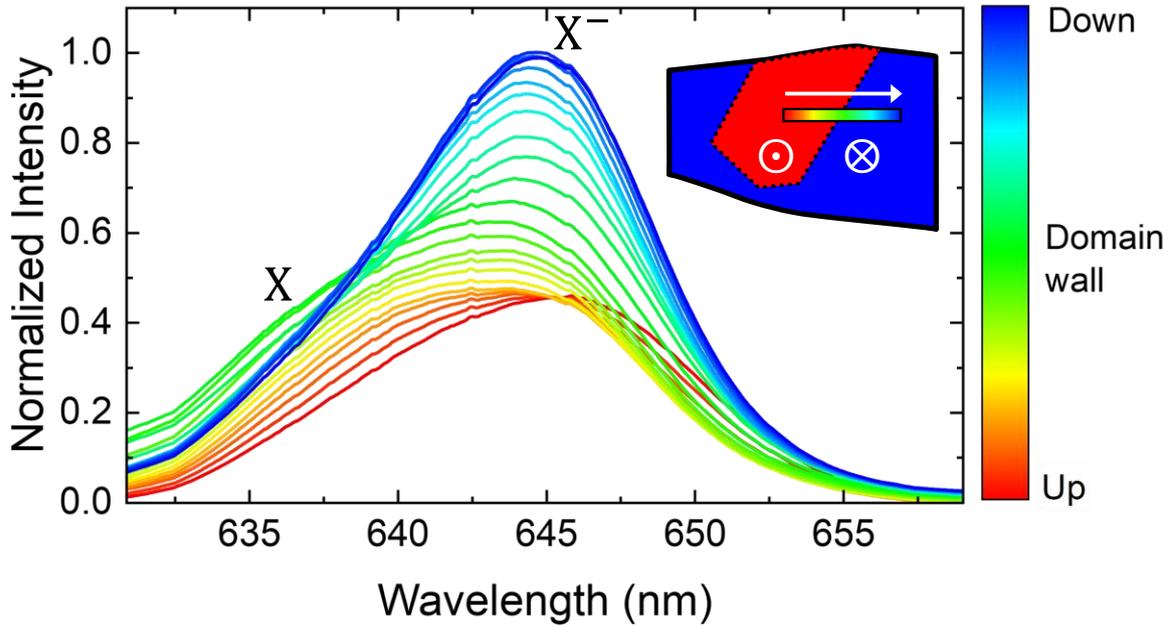

**Figure 4**. Spatial evolution of the 10 K photoluminescence of 1L-MoS$_2$ (sample S1) in the vicinity of a domain wall. The colours of the spectra refer to the position indicated in the inset. Inset: Scheme of the studied region of the sample. The Pup domain is shown in red, and the Pdown domain in blue. The white arrow indicates the scanning direction.

A deeper understanding of the influence of ferroelectric substrates in MoS$_2$ requires addressing the effects of domain walls (DWs) that separate regions with opposite polarization. DWs surfaces in LN are characterized by an intense out-of-plane electric field able to modulate the electron density in 1L-MoS$_2$ as recently demonstrated.[17] Moreover, studying DWs offers valuable insights, as they enable the formation of ultra-thin p-n junctions in MoS$_2$ without the need of light illumination.[17] Here, we extend the study to cryogenic temperatures allowing for a more precise study of exciton and trion energies and linewidths near domain walls and its effect on 1L-MoS$_2$.

**Figure 4** shows the spectra obtained by luminescence scanning microscopy at 10 K in the vicinity of a ferroelectric DW, following a linear trajectory that crosses from the P$_{up}$ domain to the P$_{down}$ domain, as schematically depicted in the inset of Figure 4. A gradual evolution of the spectra near the DW surface is observed, showing a clear increase in the excitonic contribution (green spectra).

To unravel the behaviour of both exciton and trion quasiparticles, the PL spectra were deconvoluted to analyze their intensity, peak position and linewidth in the vicinity of DWs (see **Figure 5**).

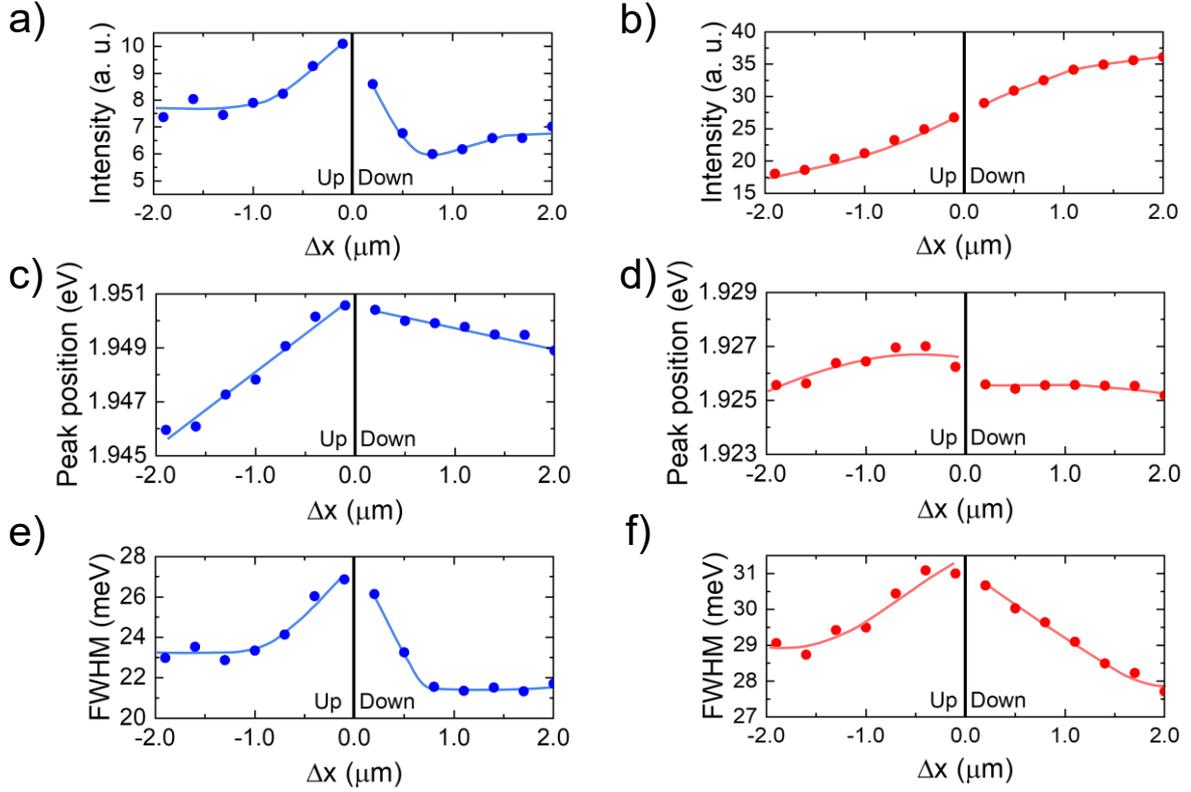

**Figure 5**. Spatial evolution of the emission features of exciton (blue) and trion (red) in the vicinity of a DW. a) and b) spatial evolution of the emission intensity. c) and d) Spatial evolution of the peak position. e) and f) spatial evolution of the linewidth. The domain wall is located at Δx = 0 µm. The results are obtained from the spectra at 10 K. Solid lines are guides for the eye.

Figures 5a and b show the emission intensities of excitons and trions, respectively, in the proximities of the DW. As observed, the exciton PL exhibits a sharp enhancement on the $P_{up}$ side of the DW followed by a clear decrease on the $P_{down}$ side. In contrast, the trion intensity increases monotonically along the scanning direction without any singularity near the DW, reaching a value on the $P_{down}$ domain that exceeds that of the $P_{up}$ domain, consistent with the enhancement of the electron density in the $P_{down}$ domain (see Figure 2c).

At this point, it is worth noting that although DWs typically span only a few nanometers, the combined effects of the strong out-of-plane electric field extending into the domain surface and the limitations in lateral spatial resolution account for the observed spatial extension around the domain wall in our experimental results.

A detail of the singularity observed in the emission intensity of exciton around the DW is shown in **Figure 6**a. According to previous results, it can be related to the presence of a strong out-of-plane electric field on the surface of DWs. As reported, this strong field abruptly changes its sign on each side of the DW, originating a highly inhomogeneous electric field distribution on the domain pattern surface.[29,30] See Figure 6b. Consequently, the observed variations in exciton intensity around the DW can be attributed to the field at DW, which influences doping in 1L-MoS$_2$. When approaching the DW from the $P_{up}$ domain, the abrupt decrease in the out-of-plane field reduces the electron doping of 1L-MoS$_2$, resulting in a pronounced increase in exciton intensity. As the field abruptly changes on the P$_{down}$ side of the DW, the electron density is significantly enhanced, which is correlated with the decrease in the exciton PL intensity. This nanometric control over charge carriers leads to the formation of deterministic p-n junctions on DWs, which, as previously shown by the authors through electrical measurements, exhibit a diode-like rectifying behavior.[17] It should be noted that while exciton photoluminescence is highly sensitive to electric fields, trion photoluminescence exhibits a weaker dependence on the doping level.[7,9] Accordingly, no abrupt changes are observed in the trion emission when crossing the domain boundary, and its intensity exhibits a monotonic evolution from one domain to the other.

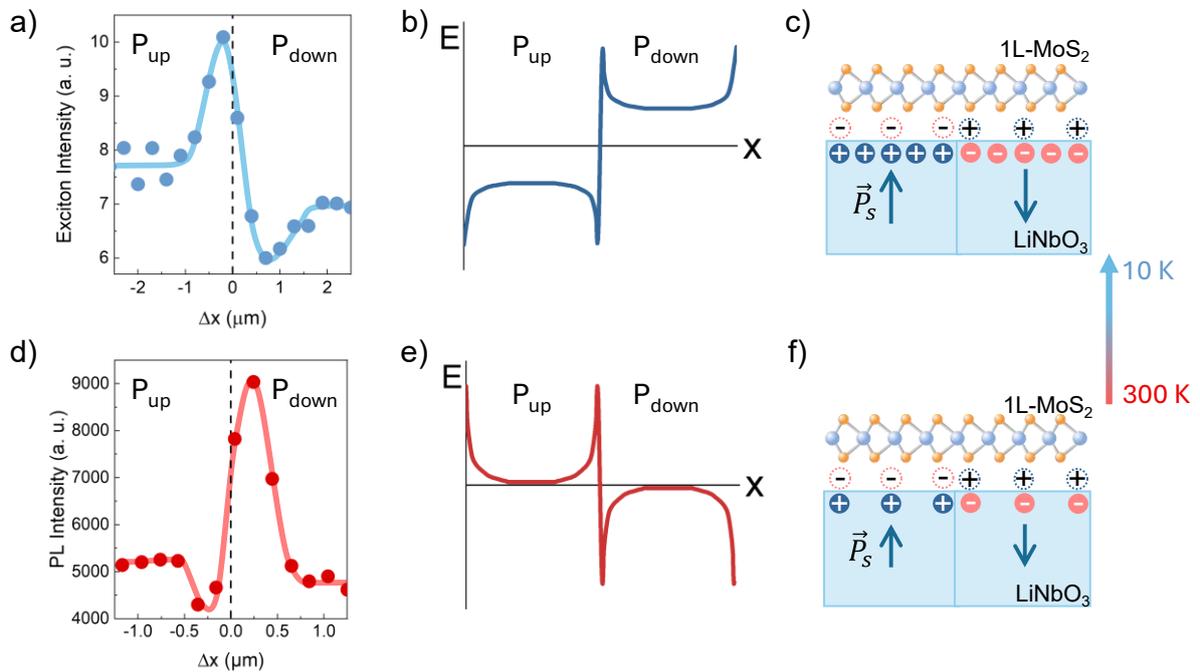

**Figure 6.** a, d) Detailed view of the spatial evolution of the exciton emission intensity at 10 K and 300 K, respectively, in the vicinity of a DW. (b, e) Schematic representation of the electric field on the surface of partially and fully screened ferroelectrics, respectively, according to refs.[29,30]. f, c) Schematics of the cross section of the PPLN. The charge balance at the 1L-MoS2 interface is illustrated. The polarization and screening charges are denoted by solid and dashed circles, respectively.

The effect of the DW of LN on the luminescence intensity of MoS$_2$ has also been observed at room temperature.[17] However, decreasing the temperature to 10 K reveals a trend opposite to that observed at room temperature. Specifically, at cryogenic temperature, the increase in the exciton emission is detected on the P$_{up}$ side of the wall, in contrast to the behavior observed at room temperature, where the sharp increase in exciton emission occurs on the P$_{down}$ domain of the wall (see Figures 6a and 6d). This change is again associated with the variation in the screening-to-polarization charge balance, which, upon cooling to cryogenic temperature, favors the presence of partially screened domain surfaces, rather than fully screened ones at room temperature.

Figures 6b and 6e schematically compare the out-of-plane field component at the surface of the alternating domain structure in LN upon cooling the system to 10 K and at 300 K, respectively. At 300 K, the spontaneous polarization of the ferroelectric LN substrate is expected to be screened by the MoS$_2$ monolayer, resulting in a highly nonuniform out-of-plane electric field distribution due to the antiparallel ferroelectric domain structure, as schematized in Figure 6e.[29,30]. In fully screened substrate, the electric field at the DWs is around 15000 V/m and abruptly changes its sign near the DW from one domain to another. However, on the domain surfaces, the field remains close to zero.[29] In contrast, under partial screening—caused by a shift in the balance between polarization and screening charges as Ps increases—the electric field reverses its sign not only on the domain surfaces, but also at the DWs.[30].

In contrast, under partial screening—caused by a shift in the balance between polarization and screening charges as $P_s$ increases—the electric field reverses its sign not only on the domain surfaces but also at the DWs.

These results corroborate the role of DWs as position-dependent, spatially defined nanometric gates in the MoS$_2$ monolayer and point out the potential of harnessing pyroelectricity to switch p-n junctions via temperature control. Specifically, the inversion of the relative doping of 1L-MoS$_2$ when the temperature is reduced from 300 K to 10 K, due to the pyroelectric effect, modifies the charge balance at the MoS$_2$/LN interface and consequently reverses the doping of the junction from p-n to n-p or vice versa.

Figures 6c and 6f also explain the difference in the electronic doping of MoS$_2$ on both domain $P_{up}$ and $P_{down}$ surfaces as the temperature decreases from 300 K to 10 K driven by pyroelectric effect. In fact, the associated increase in spontaneous polarization, $P_s$, leads to an increase in the positive or negative polarization charge, depending on the domain orientation. However, at low temperature, the variation in polarization charge at the 1L-MoS$_2$/LN interface cannot be fully compensated by a corresponding variation in the screening charge. As a result, the imbalance between screening and polarization charges generates an electric field that shifts the Fermi level relative to the valence and conduction bands, thereby explaining the doping difference on each type of domain originated by the change in $P_s$.

Finally, the variations in linewidth and energy of the exciton and trion emissions in 1L-MoS$_2$ near DWs suggest the influence of a strain-related mechanism. As shown in Figures 5c and 5d, both exciton and trion emission energies undergo a notable blueshift, around 4 meV for the exciton and 2 meV for the trion. Their linewidths exhibit a noticeable broadening around the domain wall, around 4 meV and 2.8 meV, respectively (see Figures 5e and 5f). This broadening can be attributed to the presence of strain within the monolayer on the DW.

To the best of our knowledge, no experimental studies have investigated the impact of strain on MoS$_2$ emission linewidth at low temperatures. However, previous reports at room temperature have associated emission broadening in MoS$_2$ with strain effects.[31] In fact, the strong anisotropy in the thermal expansion coefficients of the LN substrate, significantly lower along the direction of the spontaneous polarization ($4\times10^{-6}$ K$^{-1}$ at 300 K) compared to the in-plane direction ($1.5\times10^{-5}$ K$^{-1}$ at 300 K),[32] accounts for the localization of strain around the domain wall, a region of inhomogeneity. Moreover, the integrals of the thermal expansion coefficients between 10 K and 300 K are $6.33\times10^{-4}$ (along the polar direction) and $2.24\times10^{-3}$ (in-plane direction), which differ by a factor of 3.5 and corroborate the strong anisotropy of LN thermal expansion.

On the other hand, the blueshift observed in the exciton energy may also result from localized compressive strain at the domain wall. In this context, it is important to note that, due to the much larger in-plane thermal expansion coefficient of LN compared to that of 1L-MoS$_2$ (1.5×10$^{-5}$ K$^{-1}$ and 6×10$^{-6}$ K$^{-1}$, respectively)[32,33] a compressive strain is induced in the monolayer as the temperature decreases, which concentrates on the domain wall.

In fact, the strain localized at the domain wall, leading to the observed blueshift and linewidth broadening of both quasiparticles, is consistent with previous reports.[31,34] Moreover, the overall exciton blueshift of 4 meV, as shown in Figure 5c, corresponds to a compressive strain of approximately 0.08 % when applying the gauge factor from ref.[35].

## 3. Conclusions

In this work, we have investigated the impact of the pyroelectric effect on the electronic doping of monolayer MoS$_2$ deposited on a periodically poled LiNbO$_3$ (LN) substrate. Our findings provide direct evidence that pyroelectricity in LN modulates the charge carrier density in MoS$_2$. This effect is attributed to the temperature-induced variation of the spontaneous polarization (P$_s$) of LN, which modifies the balance between polarization and screening charges at the MoS$_2$/LN interface.

Through photoluminescence (PL) spectroscopy, we determined the variation in electron density from 300 K to 10 K, which correlates well with the change in the value of the spontaneous polarization of LN over that temperature range. Additionally, our experiments near ferroelectric DWs at cryogenic temperatures revealed a strong modulation of quasiparticle emissions, highlighting the role of DWs as position-dependent gate potentials. Furthermore, our results demonstrate that the pyroelectric charge modulation effectively reverses the doping of p-n junctions at DWs, converting them into n-p junctions, with potential implications for switchable devices.

Our findings highlight the interplay between pyroelectricity, ferroelectricity in shaping the electronic and optical properties of monolayer MoS$_2$, paving the way for novel strategies in ferroelectric gating engineering of 2D materials and potential switchable devices. We also stress the unique advantage of exclusively using optical probes to explore the formation of p-n MoS$_2$, providing flexibility on the characterization of monolayer based optoelectronic devices and suggesting exciting prospects for nanophotonic applications.

## 4. Experimental Section

*Domain fabrication in LiNbO$_3$ crystals*: two-dimensional patterns of hexagonal ferroelectric domains were created in a 0.5 mm-thick, z-cut monodomain LiNbO$_3$ (LN) crystal by Direct Electron Beam Writing (DEBW) with a Philips XL30 SFEG electron microscope controlled by an Elphy Raith nanolithography software. The beam current and acceleration voltage were set at 0.3 nA and 15 kV, respectively. During the irradiation process, the c+ (P$_{up}$) face was coated with a 100 nm-thick Al film acting as a ground electrode. The resulting hexagonal domains have an average diameter of 5 µm and a periodic spacing of approximately 10 µm. More details can be found elsewhere.[36] After the process, the crystal was polished up to optical quality.

*MoS$_2$ flakes exfoliation and transfer*: two different MoS$_2$ flakes, obtained from bulk crystals supplied by Manchester Nanomaterials and HQ Graphene with distinct background doping levels —in the order of $10^{13}$ e/cm$^2$ (S1) and $10^{14}$ e/cm$^2$ (S2)— were mechanically exfoliated and transferred onto the polar surface of the hexagonally poled LN substrate,[37] ensuring that the MoS$_2$ overlaps both P$_{down}$ and P$_{up}$ domains. The number of layers of the MoS$_2$ flakes was determined by employing differential micro-reflectance spectroscopy in the 400-900 nm range.[38] A commercial PPLN crystal was also used as a substrate to confirm that the behavior of the S1 monolayer agrees with previously reported results and is independent on the preparation method of the ferroelectric domain structure. For the sake of comparison, this substrate was used for the measurements of Figure 2b and 6d.

*Temperature-controlled micro-photoluminescence experiments*: micro-photoluminescence (μ-PL) experiments were carried out in a custom-made setup. A 50x objective lens (NA = 0.73) mounted on a piezoelectric platform was employed for both focusing the excitation source and collecting the emitted PL in backscattering geometry. The samples were mounted within a closed-loop optical cryostat cooled with liquid helium under vacuum conditions for the temperature-controlled μ-PL measurements. The excitation source consisted of a 532 nm laser focused to a spot size of 1.5 μm on the sample surface. The collected PL signal was filtered using a 532 nm long-pass edge filter and subsequently dispersed by a single-grating monochromator (wavelength resolution around 0.02 nm) onto a liquid nitrogen-cooled CCD detector. To achieve higher spatial resolution, room-temperature measurements in Figure 6d were taken in the absence of the cryostat, allowing the use of a 100x objective lens (NA = 0.9).

**Acknowledgements**

J.F-M, M.O.R, and L.E.B. acknowledge funding from the Spanish State Research Agency MICIU/AEI/10.13039/501100011033 under grant PID2022-137444NB-I00. G.L-P. acknowledges financial support from the Spanish State Research Agency under grant PID2022-138908NB-C32 and the the Ramón y Cajal contract RYC2023-044003-I. H.P-M. acknowledge funding from the Spanish State Research Agency under grant PID2023-148061NB-I00. The authors acknowledge AEI under grant "Maria de Maeztu" Programme for Units of Excellence in R&D CEX2023-001316-M.

**Conflict of Interest**

The authors declare no conflict of interest.

**Data Availability Statement**

The data that support the findings of this study are available from the corresponding author upon reasonable request.